# Advances in Speech Vocoding for Text-to-Speech with Continuous Parameters

Mohammed Salah Al-Radhi[1], Tamás Gábor Csapó[1,2], and Géza Németh[1]
[1] Department of Telecommunications and Media Informatics
Budapest University of Technology and Economics, Budapest, Hungary
[2] MTA-ELTE Lendület Lingual Articulation Research Group, Budapest, Hungary
{malradhi, csapot, nemeth}@tmit.bme.hu

**Abstract**
Vocoders received renewed attention as main components in statistical parametric text-to-speech (TTS) synthesis and speech transformation systems. Even though there are vocoding techniques give almost accepted synthesized speech, their high computational complexity and irregular structures are still considered challenging concerns, which yield a variety of voice quality degradation. Therefore, this paper presents new techniques in a continuous vocoder, that is all features are continuous and presents a flexible speech synthesis system. First, a new continuous noise masking based on the phase distortion is proposed to eliminate the perceptual impact of the residual noise and letting an accurate reconstruction of noise characteristics. Second, we addressed the need of neural sequence to sequence modeling approach for the task of TTS based on recurrent networks. Bidirectional long short-term memory (LSTM) and gated recurrent unit (GRU) are studied and applied to model continuous parameters for more natural-sounding like a human. The evaluation results proved that the proposed model achieves the state-of-the-art performance of the speech synthesis compared with the other traditional methods.

## 1. Introduction

Statistical parametric speech synthesis (SPSS), which generally uses a vocoder to represent speech signals, have reached a high degree of popularity over the last few years. Owing to the statistical modelling, numerous vocoders have been effectively used in text-to-speech (TTS) and speech transformation. However, its main shortcoming is the instability of the synthesis quality (Karaiskos et. al 2008). Recently, new generative models have been developed with neural network architectures, such as WaveNet (Oord et. al 2016), to synthesize speech without using a vocoder. Although WaveNet gives state-of-the-art performance and offers a natural synthesized speech, it takes a huge amount of data and computation power which causes it hard to train and apply. For that reason, vocoder-based SPSS could be offered a rapid and flexible way to expand the perceptual quality of synthetic speech (e.g. to be included in low resource devices like smartphones).

Our recent work in SPSS (Al-Radhi et. al 2017a), we developed a vocoder using continuous fundamental frequency (contF0), maximum voiced frequency (MVF), and mel-generalized cepstral (MGC), which was successfully applied in a deep neural network based on TTS (Al-Radhi et. al 2017b). Likewise other speech analysis and synthesis systems (e.g. an absence of accurate noise modeling (Degottex et. al 2018)), the noise component in continuous vocoder is not yet precisely modelled, which degrades its synthetic quality. To overcome this issue, we firstly propose a continuous noise masking (cNM) method for the purpose of improving the perceptual aspect of synthetic speech. This approach enables to mask out nearly all the noise residuals, and allows to reconstruct the voiced and unvoiced (V/UV) parts correctly. Consequently, accurate reconstruction of noise in speech frames (such as in breathiness and hoarseness) is necessary for the synthetic speech to accomplish the desired quality results

The newest results in DNN-TTS have shown that it is possible to synthesize the samples of speech directly, without using the vocoders as an intermediate step (Oord et. al 2016) (Arik et. al 2017). However, there are several drawbacks that we should take into consideration: a) it requires for each speaker a large quantity of voice data and high computation power for training model make it complicated to work in real-time applications; and b) neural models (e.g. WaveNet) are naturally serial (it needs to be repeated sequentially, one sample at a time) which cannot fully employ parallel processors (e.g. GPUs). Hence, we are aiming to propose a solution in this work to get higher-quality sound with a robust vocoder based TTS system.

## 2. Proposed Methodology

### 2.1. *Continuous Noise Masking*

The goal of this study is to eliminate the whole buzziness of the voice that is perceptible in our baseline continuous vocoder (Al-Radhi et. al 2017b). Noise masking is an essential technique to remove the noise artifacts in the time-frequency domain to increase the performance of the speech synthesizer. Recently, Degottex et. al 2018 developes a binary noise masking (bNM) in the time-frequency space with a basic measure of harmonicity. But, bNM still lacks a lowest possible of randomness in the voiced parts due to forcing values below the threshold to zero. Thus, we suggest a novel masking method named continuous noise masking (cNM) that varies from 0 to 1 (or 1 to 0) rather than a binary 0 or 1 as in the bNM, and hence keeps the quality of the voiced segments.

To determine the cNM, we must first calculate the phase distortion deviation (PDD). Here, PDD can be estimated based on (Degottex and Erro 2014). Then, cNM can be simply calculated as

$$cNM = 1 - P\acute{D}D(f) \tag{1}$$

where $P\acute{D}D$ is a regularized PDD value applying nearest-neighbor resampling approach. During the synthesis phase in continuous vocoder, the next formulas are applied to model the speech signal $s(t)$ as shown in Figure 1:

$$s(t) = \sum_{n=1}^{N} v_n(t) + u_n(t) \tag{2}$$

where $v(t)$ and $u(t)$ are the voiced and unvoiced speech components at segment $n$. Hence, for $\forall t$

$$v_n(t) = \begin{cases} v_n(t), & cNM \leq threshold \\ 0, & cNM > threshold \end{cases} \tag{3}$$

$$u_n(t) = u_n(t) * cNM(t) \tag{4}$$

To better understanding this approach, the proposed model must satisfy these conditions: If the value of the cNM approximate for the voiced frame is larger than the threshold, then this value is masked to minimize the perceptual consequence of the residual noise as may occur in the voiced parts of the cNM (lower values), while Equation 4 supervises the unvoiced frame based on the unvoiced part of the cNM (higher values). Consequently, cNM is be able to keep parts of speech component in the weak V/UV parts. Figure 2 shows an illustration of cNM estimation on a female speech sample, evaluated with the MVF contour. It can be observed that the cNM similarly tracks the real V/UV parts of the MVF. If the segment is voiced, the cNM should be lower value to dedicate that this area is voiced and must remove any other noise artifacts depends on the threshold. Whereas on the other hand, if the segment is unvoiced, the cNM should be higher value to dedicate that this area is unvoiced and must mask any other higher harmonics frequencies depends on the threshold. As a result, this method can save parts of speech components and be able to carry high-quality speech synthesis.

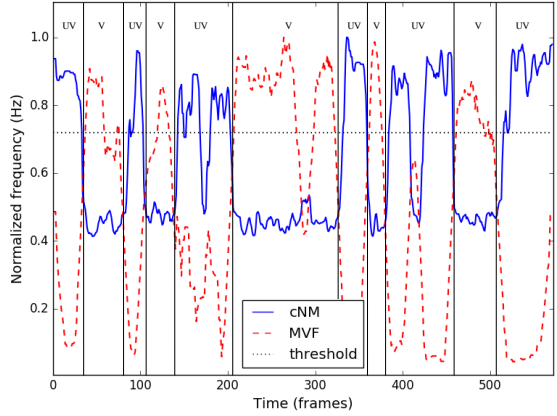

**Figure 1:** Example of the cNM (blue line) plotted across the MVF (red dashed line), where threshold = 0.77 (black dotted line). English sentence: "I was not to cry out in the face of fear." from a female speaker.

## 2.2. Training Model Based on Bi-LSTM

Bidirectional LSTM was originally proposed by Schuster and Paliwal in 1997, and it is a commonly applied for speech synthesis (Fan et. al 2014).

For a certain input variable $x = (x_1, \ldots, x_T)$, hidden state variable $h = (h_1, \ldots, h_T)$, and outputs variable $y = (y_1, \ldots, y_T)$, Bi-LSTM splits the whole state neurons in a forward state vector $\vec{h}$ (positive time direction), and backward state vector $\overleftarrow{h}$ (negative time direction). The iterative process of the suggested training model can be expressed here as

$$\vec{h}_t = tanh(W_{x\vec{h}}x_t + W_{\vec{h}\vec{h}}\vec{h}_{t-1} + b_{\vec{h}}) \tag{5}$$

$$\overleftarrow{h}_t = tanh(W_{x\overleftarrow{h}}x_t + W_{\overleftarrow{h}\overleftarrow{h}}\overleftarrow{h}_{t-1} + b_{\overleftarrow{h}}) \tag{6}$$

$$y_t = W_{\vec{h}y}\vec{h}_t + W_{\overleftarrow{h}y}\overleftarrow{h}_t + b_y \tag{7}$$

Where $W$ is the weight matrix, $b$ is the bias vectors, and $f(\cdot)$ means an activation function. Likewise to the feed forward neural network, Bi-LSTM tries to reduce the mean squared error function between the target y and the prediction $\hat{y}$ output

$$E = \frac{1}{n}\sum_{i=1}^{n}(y_i - \hat{y}_i)^2 \tag{8}$$

As a result, four feed-forward hidden layers each containing 1024 units, followed by a single Bi-LSTM layer with 385 units to train the continuous parameters.

## 3. Experimental Evaluation and Discussion

### 3.1. Datasets

A CMU-ARCTIC corpus (Kominek and Black 2003) was used to assess the sound quality of the developed algorithm. Two speakers were selected, denoted as BDL (American English, male) and SLT (American English, female). Every speaker produced 1132 sentences. In the vocoding experiments, 25 sentences from each speaker were taken randomly to be investigated and synthesized with the baseline (Al-Radhi et. al 2017b), STRAIGHT (Kawahara et. al, 1999), log domain pulse model (PML) (Degottex et. al 2018), and proposed vocoder.

### 3.2. Objective Evaluation

The empirical cumulative distribution function (Waterman and Whiteman 1978) of PDD values are computed and showed in Figure 2 to see how far these methods are from the natural signal. It can be observed that the higher mode of the distribution (positive x-axis in Figure 2) corresponding to STRAIGHT's PDMs is obviously higher than that of the original signal, whereas the PML's PDMs is lower. In contrast, the higher mode of the distribution corresponding to the proposed algorithm has good synthesized performance and almost matching the natural speech signal than others. Whereas the outcome of STRAIGHT and the baseline vocoders seem considerably worse than PML. PML's PDMs yields the second good performance on the lower mode of the distribution (negative x-axis in Figure 2) behind the proposed model. This result is possibly clarified by the fact that cNM can significantly minimize any residual buzziness.

From objective metrics (Table 1), experimental results validated that the proposed RNN models can increase the naturalness of the speech synthesized expressively over the DNN baseline. Particularly, the Bi-LSTM network gets the best performance than others. From the results, we can say that the proposed continuous vocoder within a topology of Bi-LSTM has worked satisfactory and achieved the highest naturalness scores among other neural network frameworks.

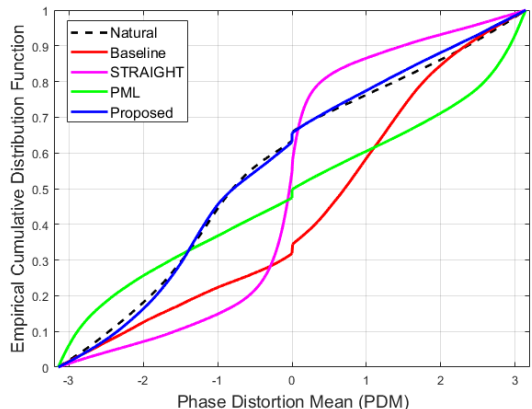

**Figure 2:** Empirical cumulative distribution function of PDMs using four vocoders compared with the natural speech signal.

### 3.3. Subjective Evaluation

In order to calculate the perceptual quality of the developed method, we performed a web-based MUSHRA (MUlti-Stimulus test with Hidden Reference and Anchor) listening experiment (ITU-R 2001). We evaluated natural sentences with the synthesized ones from the baseline, proposed, STRAIGHT, PML, and an anchor system. The participants had to assess the naturalness of each stimulus relative to the reference (which was the natural sentence), from 0 (highly unnatural) to 100 (highly natural). The listening test samples can be located online[1]. 18 participants (9 males, 9 females) with a mean age of 29 years were invited to run the online perceptual test. Figure 3 shows that the continuous vocoder gives better capability to synthesize the speech compared to the PML and STRAIGHT vocoders. Hence, cNM presents a good alternative approach to reconstruct noise than other methods (for example, bNM).

**Table 1:** Objective evaluation metrics for training models based on continuous vocoder for female and male speakers. Smaller value shows better performance excluding the CORR.

| Systems | MCD (dB) | | MVF (Hz) | | F0 (Hz) | | CORR | | Validation error | |
|---|---|---|---|---|---|---|---|---|---|---|
| | Female | Male | Female | Male | Female | Male | Female | Male | Female | Male |
| DNN | 4.923 | 4.592 | 0.044 | 0.046 | 17.569 | 22.792 | 0.727 | 0.803 | 1.543 | 1.652 |
| LSTM | 4.825 | 4.589 | 0.046 | 0.047 | 17.377 | 23.226 | 0.732 | 0.793 | 1.526 | 1.638 |
| GRU | 4.879 | 4.649 | 0.046 | 0.047 | 17.458 | 23.337 | 0.731 | 0.791 | 1.529 | 1.643 |
| Bi-LSTM | 4.717 | 4.503 | 0.042 | 0.044 | 17.109 | 22.191 | 0.746 | 0.809 | 1.517 | 1.632 |

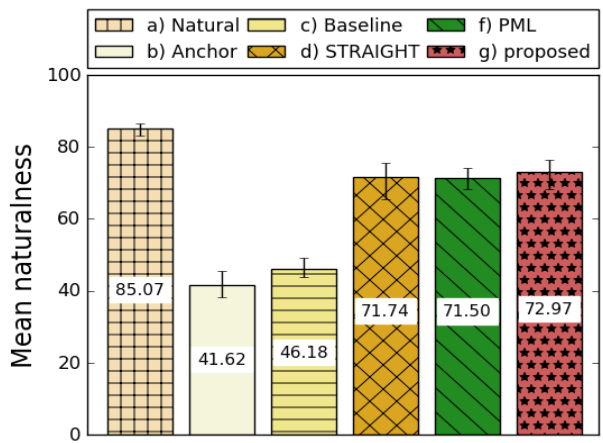

**Figure 3:** Subjective evaluation scores for the naturalness question. Error bars display the bootstrapped 95% confidence intervals.

## 4. Conclusions

This paper proposes an encouraging technique to reconstruct the noisiness of the speech signal in a continuous vocoder. We have explained an implementation of how to produce such a continuous noise masking to prevent any residual buzziness. Moreover, a Bi-LSTM based RNN was found to perform better training model with continuous vocoder than other network topologies. It was demonstrated in the experiment that the continuous vocoder gives an acceptable synthetic speech compared to the PML and STRAIGHT vocoders.

---

[1] http://smartlab.tmit.bme.hu/cNM2019

**Acknowledgements**

The research was partly supported by the European Union's Horizon 2020 research and innovation programme under grant agreement No. 825619 (AI4EU), and by the National Research Development and Innovation Office of Hungary (FK 124584 and PD 127915). The Titan X GPU used was donated by NVIDIA Corporation.